\documentclass{nature}
\usepackage[utf8]{inputenc}
\usepackage{graphicx}
\usepackage{float}
\usepackage{upgreek}
\usepackage{color,amsmath}
\usepackage[normalem]{ulem}
\usepackage{siunitx}
\usepackage{enumitem}
\usepackage{ulem}
\usepackage{amsmath,amsfonts}
\usepackage{lineno}
\usepackage{siunitx}
\usepackage{color,soul}
\usepackage{hyphenat}
\usepackage[labelfont=bf]{caption}
\usepackage{booktabs}
\usepackage{pdfpages}

\restylefloat{table}

\newcommand{\doublespacing}[1]{ \linespread{1.1}\selectfont #1}

\title{Defeating depolarizing fields with artificial flux closure in ultrathin ferroelectrics}

\author{Elzbieta Gradauskaite\textsuperscript{1,*}, Quintin N. Meier\textsuperscript{2}, Natascha Gray\textsuperscript{1}, Marco Campanini\textsuperscript{3}, Thomas Moran\textsuperscript{4}, Bryan D. Huey\textsuperscript{4}, Marta D. Rossell\textsuperscript{3}, Manfred Fiebig\textsuperscript{1}, Morgan Trassin\textsuperscript{1,*} \\ \textsuperscript{*}elzbieta.gradauskaite@mat.ethz.ch, morgan.trassin@mat.ethz.ch}

\makeatletter
\let\saved@includegraphics\includegraphics
\AtBeginDocument{\let\includegraphics\saved@includegraphics}
\renewenvironment*{figure}{\@float{figure}}{\end@float}

\makeatother

\DeclareUnicodeCharacter{2212}{-}

\begin{document}

\maketitle
\begin{affiliations}
 \item Department of Materials, ETH Zurich, 8093 Zurich, Switzerland
 \item Univ. Grenoble Alpes, CEA, LITEN, DEHT, 38000 Grenoble, France
 \item Electron Microscopy Center, Empa, 8600 Dübendorf, Switzerland
 \item Department of Materials Science and Engineering, University of Connecticut, 06269 Storrs, USA
\end{affiliations}

\textbf{Material surfaces encompass structural and chemical discontinuities that often lead to the loss of the property of interest in the so-called dead layers. It is notably problematic in nanoscale oxide electronics, where the integration of strongly correlated materials into devices is obstructed by the thickness threshold required for the emergence of their functionality\cite{Junquera2003,Fong2004a}. Here, we report the stabilization of ultrathin out-of-plane ferroelectricity in oxide heterostructures through the design of an artificial flux-closure\cite{McQuaid2011,Tang2015} architecture. Inserting an in-plane polarized ferroelectric epitaxial buffer provides continuity of polarization at the interface, and despite its insulating nature we observe the emergence of polarization in our out-of-plane-polarized model ferroelectric BaTiO\textsubscript{3} from the very first unit cell. In BiFeO\textsubscript{3}, the flux-closure approach stabilizes a conceptually novel 251\textdegree{} domain wall. Its unusual chirality is likely associated with the ferroelectric analog to the Dzyaloshinskii–Moriya interaction\cite{Zhao2021}. We thus see that in an adaptively engineered geometry, the depolarizing-field-screening properties of an insulator can even surpass those of a metal and be a source of new functionalities\cite{Chauleau2020}. This should be a useful insight on the road towards the next generation of ferroelectric-based oxide electronics\cite{Garcia2014}.}

Uncompensated bound charges at the surfaces of ferroelectric films trigger a depolarizing field\cite{Mehta1973}. It is oriented opposite to the spontaneous polarization and highest when this polarization is normal to the surface---the preferred orientation for ferroelectric devices. Thus, the ferroelectric properties are attenuated with decreasing thickness and disappear completely below about 5 unit cells (u.c.) in most perovskites\cite{Fong2004a,DeLuca2017a}. This holds true even when ferroelectrics are deposited on metallic electrodes, which all have a finite screening length and normally cannot fully compensate for the bound charges\cite{Junquera2003}. To maintain polarization in the ultrathin regime, attempts have been made to minimize the surface-charge accumulation or to enforce polar displacements from the ferroelectric\(|\)electrode interface via interface engineering. For instance, interface chemistry can be tailored to create a favorable charge-screening environment and hence a net polarization in the ultrathin regime\cite{Strkalj2020}. Polar metals were also suggested to eliminate the critical thickness in ferroelectrics by inducing inversion symmetry-breaking from the bottom interface\cite{Puggioni2018a}, but their experimental implementation remains nevertheless elusive due to the scarcity of lattice-matching systems.

The interfacial polarization discontinuity can be avoided with the use of in-plane polarized ferroelectrics, which are less susceptible to the depolarizing field even at low thicknesses\cite{Keeney2020,Gradauskaite2020a}. The in-plane polarization anisotropy, however, is incompatible with the coveted state of the art, energy-efficient out-of-plane capacitor device geometry. Thus, the mutually exclusive benefits of out-of-plane and in-plane polarized ferroelectrics pose a serious obstacle to the ongoing quest for the next generation of oxide electronics. 

Here we combine the ``best of both worlds'' by eliminating the interfacial polar discontinuity in an out-of-plane-polarized ferroelectric heterostructure by creating a flux-closure-like domain architecture. We accomplish this by interfacing two of the most debated out-of-plane polarized systems, BaTiO\textsubscript{3} (BTO) and BiFeO\textsubscript{3} (BFO), with an in-plane polarized, layered ferroelectric of the Aurivillius phase\cite{aurivilliusmain} Bi\textsubscript{5}FeTi\textsubscript{3}O\textsubscript{15} (BFTO). In typical perovskite heterostructures, the substrate lattice sets the uniform polarization anisotropy of the epitaxial heterostructure. In our case, the integration of the in-plane polarized BFTO with the out-of-plane polarized systems provides polarization continuity at the interface and enables the onset of an out-of-plane ferroelectric polarization from the very first u.c. The functional impact of the flux-closure-like ferroelectric interface is highlighted in two ways. First, we demonstrate local switching of the BTO polarization, despite the absence of conducting bottom electrode. And second, we stabilize polar Néel domain walls with a deterministic chirality in multiferroic BFO thin films. Thus, we introduce ferroelectric heterostructures with perpendicular in-plane and out-of-plane polar anisotropies for the design of electric polarization at the nanoscale, offering a superior alternative to metal-based ferroelectric device paradigms.

\begin{figure}[!htb] \centering
\includegraphics[width=16.5cm]{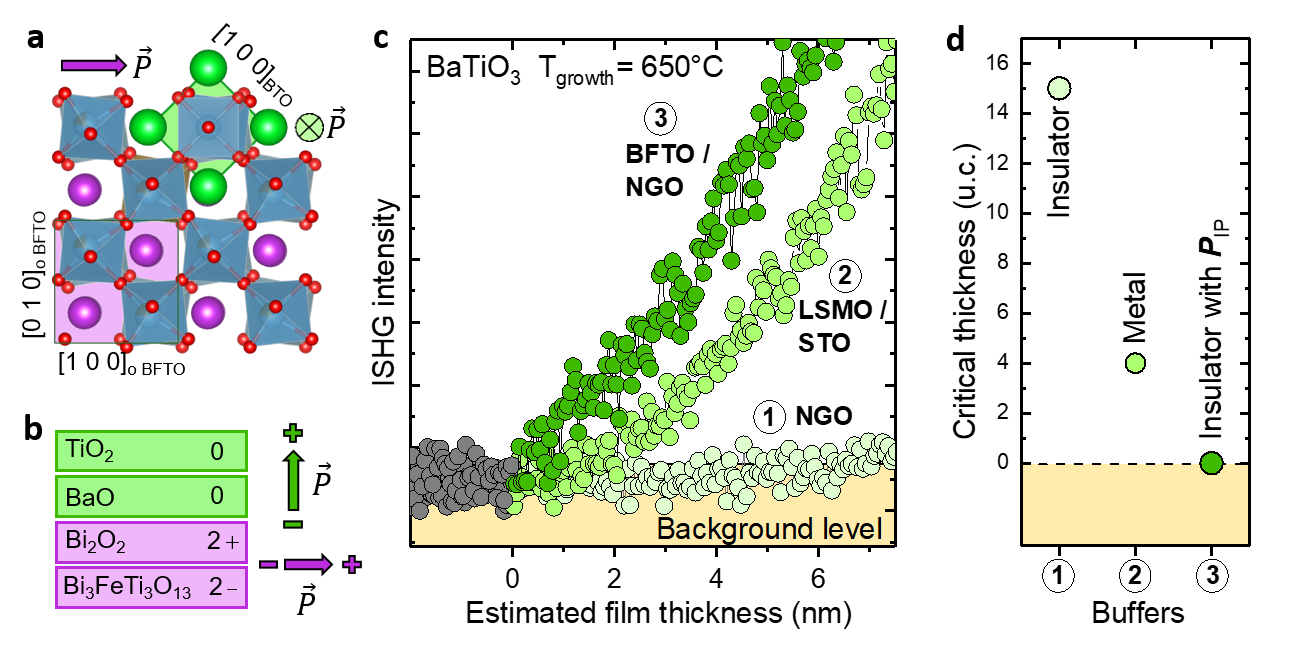}
 \doublespacing{
  \caption*{\textbf{Fig. 1 \(|\) Absence of critical thickness for ferroelectricity in BTO grown on 0.5 u.c. BFTO.}
  \textbf{a}, A visualization of epitaxial relationship between BFTO and BTO unit cells in the plane of the films. \textbf{b}, Atomic plane stacking at the BTO\(|\)BFTO interface, showing the BTO polarization direction set by the net positive layer charge at the  interface. \textbf{c}, ISHG signal tracking the BTO thin-film polarization during the identical deposition on: 1) insulating NGO (001)\textsubscript{o} substrate, 2) on LSMO electrode on STO (001) and 3) on 0.5 u.c. BFTO on NGO (001)\textsubscript{o}. \textbf{d}, Variation in critical thickness for ferroelectricity in BTO thin films depending on the buffer utilized.}}
\end{figure}

We begin our investigation by monitoring the emergence of polarization in a BTO film epitaxially grown on an in-plane-polarized BFTO buffer layer using optical second harmonic generation (SHG). This process is sensitive to inversion symmetry breaking and therefore an ideal probe for ferroelectricity. When measured during the pulsed-laser-deposition growth process, as in-situ SHG (ISHG\cite{DeLuca2017a}, see Methods), one directly accesses the emergence of polarization in ultrathin films with unit-cell accuracy thanks to the simultaneous reflection high-energy electron diffraction (RHEED) monitoring. We select the Aurivillius compound BFTO as our model in-plane-polarized ferroelectric buffer, because it can be grown with single-crystal quality on NdGaO\textsubscript{3} (NGO) (001)\textsubscript{o}\cite{Gradauskaite2020a,Gradauskaite2021a} (``o''  denotes orthorhombic indices) and exhibits a robust in-plane polarization from the first half-unit cell\cite{Gradauskaite2020a}. Furthermore, it is structurally compatible with functional perovskite oxides (Fig. 1a), and its charged layered architecture favors an upwards-pointing polarization in the BTO layer grown on top\cite{Spaldin2021,DeLuca2017a} (Fig. 1b).

The absence of an ISHG signal and, hence, of a spontaneous polarization, during the early stage of the growth of (001)-oriented BTO directly on the insulating NGO substrate, in Figure 1c, confirms the dominant role of the depolarizing field in a non-charge-screened environment. Only after passing the thickness of 15 u.c., we detect a signal corresponding to a net out-of-plane-oriented polarization. The conventional approach to combat the depolarizing field and reduce the critical thickness involves the use of a lattice-matching conducting buffer such as La\textsubscript{0.7}Sr\textsubscript{0.3}MnO\textsubscript{3} (LSMO). Indeed, the associated conduction leads to a significant improvement of the charge-screening environment. The ISHG measurement in Figure 1c reveals a reduction of the thickness threshold for the onset of ferroelectricity down to 4 u.c., however, it does not completely eradicate it. We therefore replace the metal by 0.5 u.c. of ferroelectric in-plane-polarized BFTO. Note that the first half-unit-cell of BFTO ($\sim$2 nm in height) already contains four perovskite layers, which give rise to the net in-plane polarization, and thus, exhibits the full functionality of the Aurivillius compound\cite{Gradauskaite2020a,Keeney2020a}. At first glance, this seems counterproductive because of the insulating nature of this polar buffer. However, Figure 1c shows that we obtain an ISHG signal with the very first BTO u.c. deposited on top of the BFTO; the film grows right in the ferroelectric phase, without any critical thickness threshold. The subsequent continuous ISHG increase with the ongoing BTO deposition is consistent with an out-of-plane-polarized ferroelectric single-domain state\cite{DeLuca2017a}, characteristic of BTO. Ex-situ X-ray diffraction (XRD) studies confirm the presence of (001)-oriented BTO with the expected epitaxial relationship, involving a 45\textdegree{} in-plane rotation of the BTO u.c. on the orthorhombic u.c. of the BFTO buffer (Supplementary Fig. S1, S2). We thus see that an in-plane-polarized insulating buffer layer, can surpass metallic electrodes in stabilizing an out-of-plane-polarized ferroelectric state in the ultrathin regime (Fig. 1d).

\begin{figure}[htb!] 
\centering
\includegraphics[width=16cm]{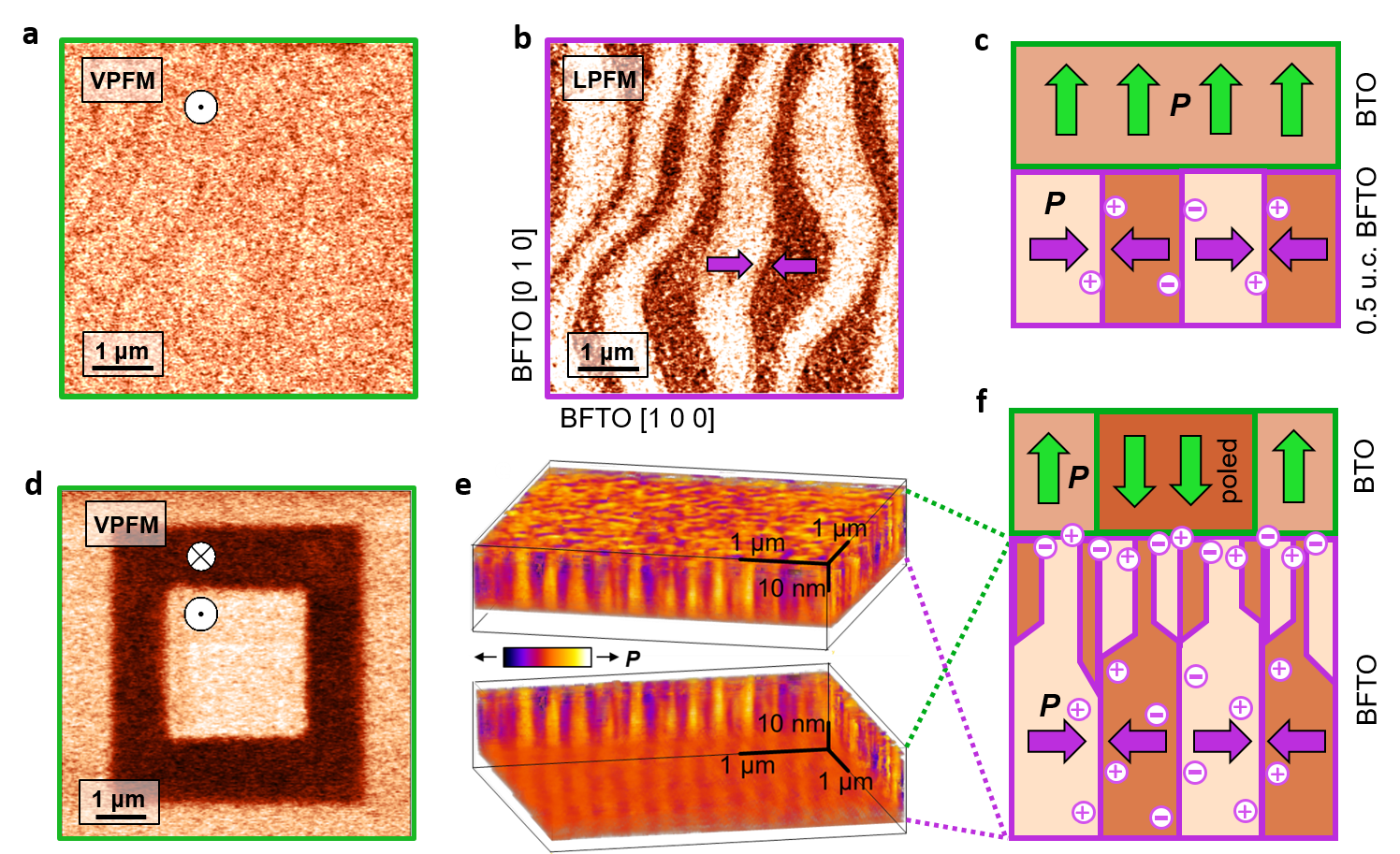}
  \doublespacing{\caption*{\textbf{Fig. 2 \(|\) Ferroelectric domain configuration in BTO\(|\)BFTO heterostructure.}
  \textbf{a}, VPFM and \textbf{b}, LPFM scans of BTO\(|\)BFTO bilayer reveal a monotonous out-of-plane polarization in BTO and a stripe pattern of in-plane-polarized domains in BFTO, respectively. \textbf{c}, Sketch of the resulting domain configuration the two layers form, comparable to partial flux-closure domains. \textbf{d}, VPFM scan confirming a successful local polarization reversal in BTO grown on 3.5 u.c. of BFTO without a metal electrode. \textbf{e}, Three-dimensional PFM-tomographic reconstructions of in-plane-polarized domains in a 5 u.c.-thick BFTO film, viewed obliquely from above and below, directly revealing distinct stripe and columnar domain regimes. \textbf{f}, Representation of the thickness-dependent domain architecture in BFTO. Higher thickness results in increased density of mobile charges at the interface, which enables the polarization switching.}}
\end{figure}

In order to shed light on this seemingly counterintuitive observation, we investigate the ferroelectric domain configuration in our BTO\(|\)BFTO bilayer by PFM. The uniform vertical-PFM (VPFM) contrast in Figure 2a is consistent with an out-of-plane-polarized single domain configuration in the top BTO layer as inferred from both ISHG and XRD. The lateral-PFM (LPFM) scan in Figure 2b, however, reveals a 200-nm-wide stripe-domain pattern with alternating tail-to-tail (TT) and head-to-head (HH) charged domain walls (CDWs) characteristic of the buried, fully in-plane-polarized BFTO film of 0.5 u.c.\cite{Gradauskaite2020a,Keeney2020a}. Thus, despite their rigid epitaxial relationship, the BTO and BFTO in the bilayer adhere to their respective polar anisotropies. The domain configuration of the BTO\(|\)BFTO heterostructure thus resembles a partial flux closure as sketched in Figure 2c. In ferroic materials flux-closure domain formation efficiently reduces the field energy by forming closed loops of field lines within the material\cite{Kittel}. Specifically, in ferroelectrics, flux closure\cite{McQuaid2011,Tang2015} effectively minimizes the polarization discontinuities and bound charge accumulation at the domain boundaries or interfaces. In our case it means that the depolarizing field acting on BTO is successfully defeated, allowing the out-of-plane polarization to establish from the very first u.c.

Remarkably, we observe that the polarization in BTO can be reversibly poled with a scanning probe tip ($\pm$5V) once the thickness of BFTO buffer increases above 2 u.c. This is rather unexpected since ferroelectric poling ordinarily requires the insertion of a conducting electrode. In order to investigate the role of buffer thickness in the BTO poling behaviour, we perform tomographic PFM\cite{Steffes2019} (see Methods) on a buffer layer of 5 u.c. (approx. 25 nm). Figure 2e shows a three-dimensional reconstruction of the domain configuration throughout the thickness of the BFTO film, with views from both above and below. The in-plane PFM contrast clearly reveals that the stripe architecture of in-plane-polarized 180\textdegree{} domains nearest the substrate gradually transitions into fine columnar domains with increasing film thickness (see Supplementary Fig. S3.1, S3.2). We conclude that the increased population of percolating CDWs, and of the corresponding density of mobile charges in buffer layers exhibiting such columnar domains (as schematically introduced in Figure 2f), emulates the functionality of a counterelectrode and thereby uniquely enables polarization switching of the BTO.

So far we considered an inherently out-of-plane-polarized ferroelectric (BTO) stabilized as ultrathin film by an in-plane-polarized buffer layer (BFTO). Replacing BTO with a ferroelectric that has its own in-plane polarization components, could bring additional degrees of freedom to the ferroelectric domain and domain-wall engineering in our flux-closing system. To explore this, we grow (001)-oriented BFO, the only known robust room-temperature multiferroic magnetoelectric material, onto the BFTO buffer. Its eight possible polarization components along pseudocubic $\langle$111$\rangle$\textsubscript{pc} directions\cite{Streiffer1998} lead to a rich variety of possible ferroelectric domain and domain-wall configurations with out-of-plane and also the desired in-plane polarization components.

\begin{figure}[!htb] \centering
\includegraphics[width = \linewidth]{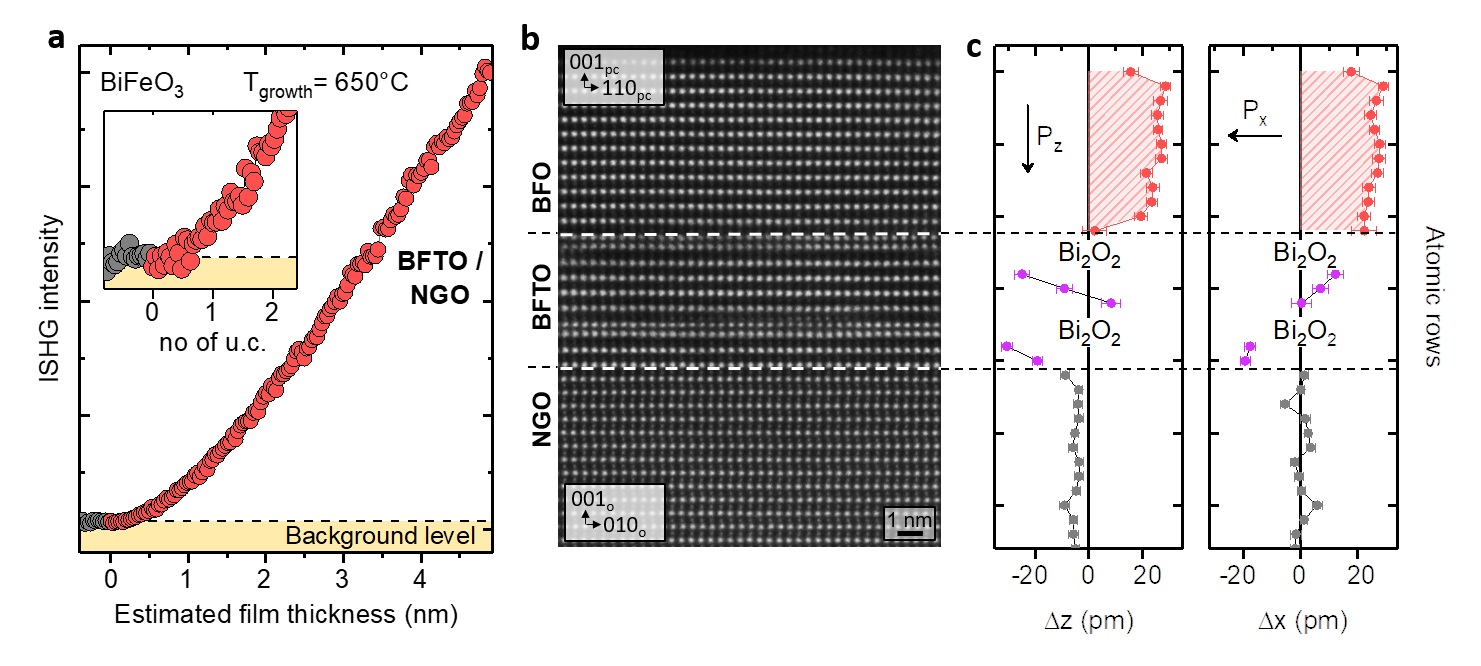}
  \doublespacing{\caption*{\textbf{Fig. 3 \(|\) Polarization continuity at the BFO\(|\)BFTO interface.} \textbf{a}, ISHG signal tracking the net out-of-plane polarization in the BFO thin film during its deposition on 1 u.c. of BFTO (red symbols). Inset shows the signal collected during the deposition of the first two u.c. \textbf{b}, HAADF-STEM image of the coherently strained BFO\(|\)BFTO\(|\)NGO heterostructure. \textbf{c}, Out-of-plane ($\Delta$z) and in-plane ($\Delta$x) polar displacements of the B-site atomic columns from the center of their two nearest A-site neighbors measured from the atomically-resolved HAADF-STEM image in panel \textbf{b}, confirming the presence of the BFO polarization form the first u.c.}}
\end{figure}

We monitor the ISHG signal during the BFO deposition onto a BFTO layer of a single u.c. As in the case of BTO, the ISHG yield (Fig. 3a) shows an onset of ferroelectricity in the BFO film with the first u.c., thus demonstrating the material-independent potential of the flux-closure-like architecture in defeating depolarizing fields. Post-deposition STEM imaging in Figure 3b reveals the high crystalline quality of the BFO\(|\)BFTO\(|\)NGO heterostructure. An evaluation of the dipolar displacements (Fig. 3c) extracted from the atomically-resolved HAADF-STEM image in Figure 3b shows that there is a clear polarization continuity at the BFO\(|\)BFTO interface, highlighting the absence of a dead layer (maps of the in-plane and out-of-plane displacements are shown in Supplementary Fig. S4).

To uncover how the interfacial flux closure affects the domain pattern and the domain wall architecture in the BFO film, we opt for high-resolution scanning-probe measurements for which a fully-strained conducting LSMO layer is inserted underneath the flux-closing BFO\(|\)BFTO heterostructure. The charge screening provided by LSMO prevents the formation of domains in the Aurivillius buffer (Fig. 4a), thus simplifying the PFM investigations while maintaining the respective polar anisotropies in the BFO\(|\)BFTO bilayer and, hence, its flux-closing properties. The two LPFM contrast levels (Fig. 4b) suggest two in-plane polarization components in the BFO film. As Supplementary Fig. S5.1 suggests, these are parallel to the uniaxial polarization axis of the BFTO. The VPFM signal is largely uniform (Fig. 4c), aside from some linear and dot-like discontinuities, which will be discussed later. This is consistent with a uniform downwards polarization across the BFO film, dictated by the BFTO atomic-plane termination\cite{Spaldin2021}. Therefore, out of the eight possible domain states, the pristine BFO in our heterostructure exhibits only two, namely those with a polarization pointing along [$11\bar{1}$]\textsubscript{pc BFO} or [$\bar{1}\bar{1}\bar{1}$]\textsubscript{pc BFO}, henceforth denominated as \textbf{P}\textsubscript{1} and \textbf{P}\textsubscript{2}, respectively (Fig. 4d). This geometry results in the formation of in-plane 109\textdegree{} domain walls, in which, unlike in commonly observed 109\textdegree{} domain walls in BFO films, the out-of-plane polarization component remains downwards oriented from one domain to the other, as sketched in Figure 4d. A restriction to such in-plane 109\textdegree{} domain walls only has never been observed in BFO films and shows that the in-plane polarized buffer can act as a powerful tool in setting the domain configuration.  
\begin{figure}[!htb] \centering
\includegraphics[width =\linewidth]{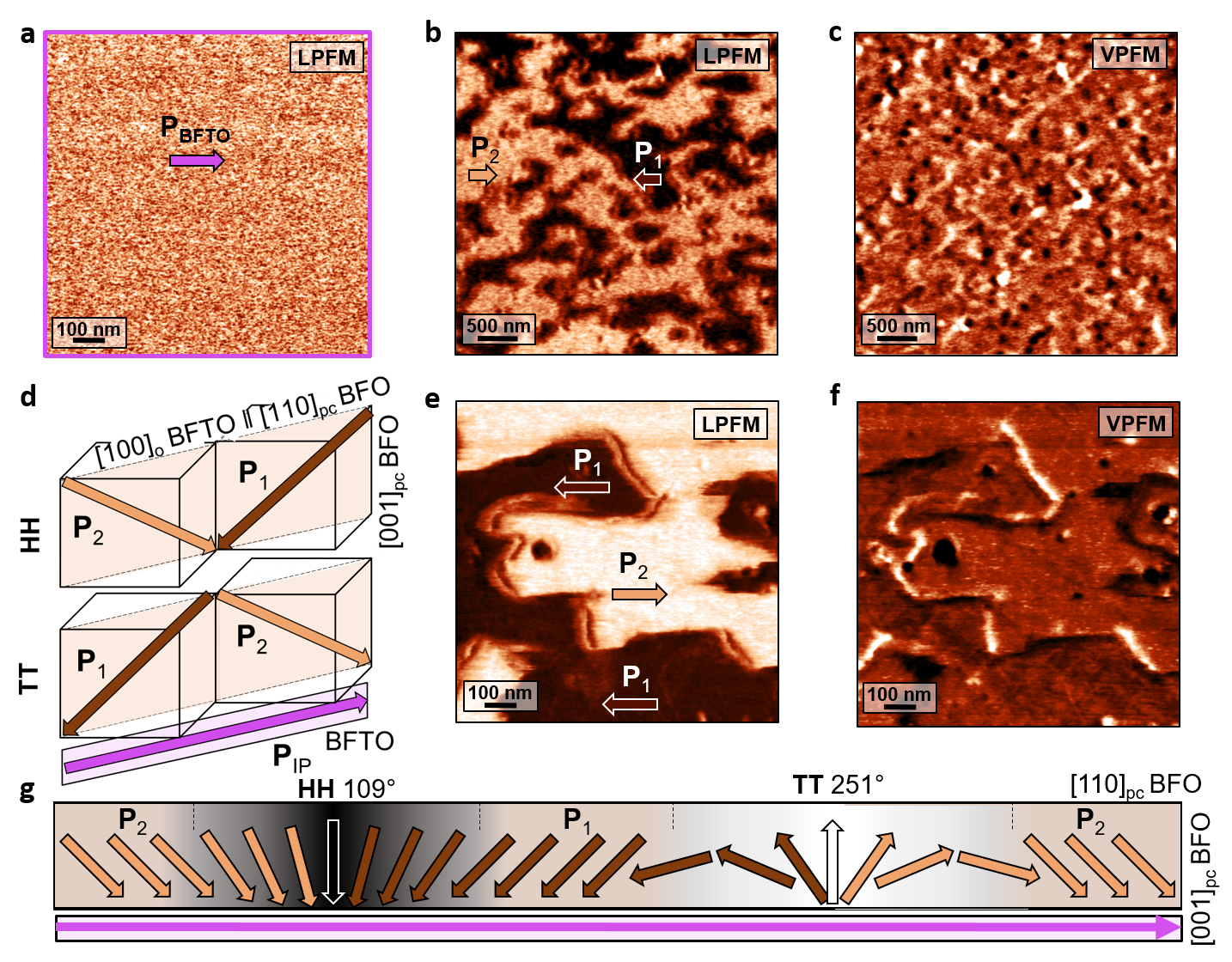}
  \doublespacing{\caption*{\textbf{Fig. 4 \(|\) Polar homochiral Néel domain walls in BFO grown on BFTO.}
  \textbf{a}, LPFM scan performed prior to the BFO deposition shows that the BFTO film exhibits a unidirectional polarization pointing along [110]\textsubscript{pc BFO}. \textbf{b}, LPFM scan reveals only [$11\bar{1}$] and [$\bar{1}\bar{1}\bar{1}$] as two in-plane polarization components of BFO when deposited on the BFTO. \textbf{c}, VPFM scan shows a downwards polarization across the film with scattered local out-of-plane-polarized features (see \textbf{e},\textbf{f}). \textbf{d},
  Schematic of in-plane 109\textdegree{} HH and TT CDWs stabilized by the in-plane-polarized buffer. \textbf{e},\textbf{f}, High-magnification LPFM (\textbf{e}) and VPFM (\textbf{f}) scans show that additional out-of-plane polarization components appear at the charged sections of domain walls, exhibiting the inhomogeneities seen in \textbf{c}. \textbf{g}, Polarization rotation profiles in the (110)\textsubscript{pc} plane across HH and TT domain walls, corresponding to downwards and upwards pointing polarization rotation, respectively. Deterministic polarization rotation at the walls implies a net chirality in the BFO film likely arising due to the unidirectional BFTO polarization underneath.}}
\end{figure}

To our surprise, we observe that the polarization is reorienting with the same sense of rotation across each of the BFO charged domain wall sections separating the 109\textdegree{} domains of the BFO. All the TT walls exhibit a rotation through an upwards polarized state (bright VPFM signal), whereas it is a downwards polarized state for the HH walls (dark VPFM signal) (see Supplementary Fig. S5.2). The reported behaviour was verified on six samples. The most striking aspect of this observation is that in the \textbf{P}\textsubscript{1}-\textbf{P}\textsubscript{2} domain wall on the right hand side of Figure 4g the dipolar reorientation prefers a 251\textdegree{} ``detour'' over the expected 109\textdegree{} rotation. In order to highlight the unusual structure of this domain wall and to distinguish it from the well-known appearance of 109\textdegree{} walls in thin films, we propose to introduce the notion of a 251\textdegree{} wall in this case. Note that the association of the HH walls to the 109\textdegree{} rotation and the TT walls to the 251\textdegree{} rotation leads to a non-zero net chirality of ferroelectric domain walls in our heterostructure, that a BFO film deposited directly onto the LSMO does not exhibit (Supplementary Fig. S6).

The homochirality of the domain walls and the fact that a 251\textdegree{} wall is usually energetically more costly than a 109\textdegree{} wall show that the BFTO buffer exerts a pronounced symmetry breaking effect. Its manifestation is reminiscent of the domain wall homochirality observed in ferromagnetic heterostructures, where it is a consequence of the Dzyaloshinskii-Moriya interaction (DMI)\cite{Emori2013,Velez2019}. It can arise from symmetry breaking at the magnetic surfaces or interfaces and generates noncollinear spin structures. Here, in close analogy, an interface with an in-plane-polarized BFTO layer could break the symmetry and give rise to the noncollinear polar textures we observe at the BFO domain walls. Mirroring the free energy invariant describing the DMI in chiral magnets and antiferromagnets\cite{Dzyaloshinskii1964,Bogdanov2002,LandauLifshitz}, the corresponding DMI-like\cite{Zhao2021} invariant for the polarization can be expressed as:
$E_{DM}\propto (P_{x}\cdot\partial_x P_{z}-P_z\cdot\partial_x P_{x})$, where $x$ corresponds to the in-plane polarization axis, and $z$ points out of plane. As for chiral magnets, it sets a deterministic rotation sense of the order parameter at the domain walls, as we demonstrate using phase field simulations (see Supplementary Fig. S7). Such ferroelectric DMI-like interactions have so far mostly been discussed from a theoretical standpoint\cite{Erb2020,Zhao2021} and the emergence of 251\textdegree{} domain walls in our BFO films is likely one of the first experimental observations of it. All this suggests the use of an in-plane polarized buffer as a novel, unforeseen route for the stabilization of polar homochirality\cite{Chauleau2020,Fusil2022}. 

Thus, our work, introduces epitaxial flux-closing heterostructures that stabilize an out-of-plane polarization in ultrathin BTO and BFO right from the very first u.c. The interfacial symmetry breaking, occurring between in-plane and out-of-plane-polarized layers, can give rise to a net polar chirality at the domain walls of the BFO film. In particular, it leads to 251\textdegree{} BFO domain walls in our heterostructures, which may be a signature of a DMI-like behavior in ferroelectrics. These results demonstrate that polar insulators can be more effective than metals in screening the detrimental depolarizing field in the ultrathin regime and can be used to initiate entirely novel functional architectures. Both aspects could make the epitaxial combination of perpendicular polar anisotropies a powerful tool in oxide-electronics research. 

\section*{\large{Methods}}

\subsection{Heterostructure Growth}
The thin films and heterostructures were grown on NGO (001)\textsubscript{o} substrates by pulsed laser deposition using a KrF excimer laser at 248 nm. The laser fluence, repetition rate, substrate temperature, and growth pressure set for individual layers are as follows: BFTO: 0.9 J cm\textsuperscript{−2}, 2 Hz, 650 \textdegree{}C, 0.075 mbar O\textsubscript{2}; BTO: 0.9 J cm\textsuperscript{−2}, 2 Hz, 650 \textdegree{}C, 0.015 mbar O\textsubscript{2}; BFO: 0.7 J cm\textsuperscript{−2}, 8 Hz, 630 \textdegree{}C, 0.125 mbar O\textsubscript{2}; LSMO: 0.9 J cm\textsuperscript{−2}, 1 Hz, 650 \textdegree{}C, 0.035 mbar O\textsubscript{2}. The thickness of the thin films was monitored using a combination of RHEED during growth and X-ray reflectivity ex-situ.

\subsection{ISHG monitoring} Optical SHG was probed in reflection and in situ, in the pulsed laser deposition growth chamber\cite{DeLuca2017a}. The probe beam of \SI{860}{\nano\metre} (ISHG of BFO films) and \SI{1200}{\nano\metre} (ISHG of BTO films) was incident onto the sample with a pulse energy of \SI{20}{\micro\joule} and onto a spot size of \SI{250}{\micro\metre} in diameter. The ISHG signal was detected using a monochromator and a photomultiplier system\cite{DeLuca2017a}. 

\subsection{PFM} The scanning probe microscopy measurements were recorded using a NT-MDT NTEGRA scanning probe microscope and two external SR830 lock-in detectors (Stanford Research) for simultaneous acquisition of in-plane and out-of-plane piezoresponse. The data acquisition was performed using a 2.3V AC modulation at 70 kHz applied to the Pt-coated tip. The ferroelectric domain configurations of the pristine BFO sample was identified using vector PFM (see Supplementary Figure S5.1). Deflection and torsion modes were recorded when measuring with cantilever perpendicular to the uniaxial polarization axis of BFTO/BFO. 

\subsection{PFM tomography} Tomographic AFM is based on sequential AFM imaging and probe-based nanomechanical milling\cite{Song2020}. Operating in conventional PFM mode\cite{Steffes2019}, specifically while biasing the conducting tip with an AC field oscillating at the torsional contact-resonance, thereby volumetrically maps the in-plane domain contrast. A PFM tomogram constructed from a sequence of 95 images, all in the same field of view but for gradually diminishing film thicknesses, is presented in Figure 2e, which overall comprises more than 1 million 15.7$\times$15.7$\times$1 nm\textsuperscript{3} voxels of local piezoresponse signals. The color contrast depicts the product of the measured amplitude and the sign of the phase for in-plane torsion of the cantilever, i.e. the lateral piezoresponse. The AFM (Asylum Research Cypher VRS) is operated in contact mode using doped diamond probes (AppNano DD-ACTA). A setpoint of approximately 2.24 µN is sufficient to mill this specimen, while torsional PFM imaging is performed with a 1-V peak-to-peak AC bias at a frequency of approximately 2.8 MHz. The amplitude and phase of the locally vibrated lever are simultaneously acquired via a lock-in amplifier (Zurich Instruments MFLI).

\subsection{STEM} Cross-sectional specimens were prepared by focused ion beam (FIB) milling with a FEI Helios NanoLab 600i. High-angle annular dark-field scanning transmission electron microscopy (HAADF-STEM) imaging was attained with a FEI Titan Themis microscope equipped with a spherical-aberration probe corrector (CEOS DCOR) operated at 300 kV. A probe semi-convergence angle of 18 mrad was set in combination with an annular semi-detection range of the HAADF detector of 66-200 mrad. The HAADF-STEM images were obtained by averaging time series of 10 frames and the atomic column positions in the averaged images were fitted with picometer precision as described in refs. \cite{Yankovich2014}, \cite{Campanini2018}. The $\Delta$z and $\Delta$x polar displacements of the B-site atomic columns in the image plane from the center of their two nearest A-site neighbors are respectively along the [100]\textsubscript{pc} and [011]\textsubscript{pc} directions.

\section*{\large{References}}

\bibliographystyle{naturemag_new.bst}
\bibliography{Aurivillius}

\begin{thebibliography}{10}
\expandafter\ifx\csname url\endcsname\relax
  \def\url#1{\texttt{#1}}\fi
\expandafter\ifx\csname urlprefix\endcsname\relax\def\urlprefix{URL }\fi
\providecommand{\bibinfo}[2]{#2}
\providecommand{\eprint}[2][]{\url{#2}}

\bibitem{Junquera2003}
\bibinfo{author}{Junquera, J.} \& \bibinfo{author}{Ghosez, P.}
\newblock \bibinfo{title}{{Critical thickness for ferroelectricity in
  perovskite ultrathin films}}.
\newblock \emph{\bibinfo{journal}{Nature}} \textbf{\bibinfo{volume}{422}},
  \bibinfo{pages}{506--509} (\bibinfo{year}{2003}).

\bibitem{Fong2004a}
\bibinfo{author}{Fong, D.~D.} \emph{et~al.}
\newblock \bibinfo{title}{{Ferroelectricity in ultrathin perovskite films}}.
\newblock \emph{\bibinfo{journal}{Science}} \textbf{\bibinfo{volume}{304}},
  \bibinfo{pages}{1650--1653} (\bibinfo{year}{2004}).

\bibitem{McQuaid2011}
\bibinfo{author}{McQuaid, R.~G.}, \bibinfo{author}{McGilly, L.~J.},
  \bibinfo{author}{Sharma, P.}, \bibinfo{author}{Gruverman, A.} \&
  \bibinfo{author}{Gregg, J.~M.}
\newblock \bibinfo{title}{{Mesoscale flux-closure domain formation in
  single-crystal BaTiO\textsubscript{3}}}.
\newblock \emph{\bibinfo{journal}{Nature Communications}}
  \textbf{\bibinfo{volume}{2}}, \bibinfo{pages}{404} (\bibinfo{year}{2011}).

\bibitem{Tang2015}
\bibinfo{author}{Tang, Y.~L.} \emph{et~al.}
\newblock \bibinfo{title}{{Observation of a periodic array of flux-closure
  quadrants in strained ferroelectric PbTiO\textsubscript{3} films}}.
\newblock \emph{\bibinfo{journal}{Science}} \textbf{\bibinfo{volume}{348}},
  \bibinfo{pages}{547--551} (\bibinfo{year}{2015}).

\bibitem{Zhao2021}
\bibinfo{author}{Zhao, H.~J.}, \bibinfo{author}{Chen, P.},
  \bibinfo{author}{Prosandeev, S.}, \bibinfo{author}{Artyukhin, S.} \&
  \bibinfo{author}{Bellaiche, L.}
\newblock \bibinfo{title}{{Dzyaloshinskii–Moriya-like interaction in
  ferroelectrics and antiferroelectrics}}.
\newblock \emph{\bibinfo{journal}{Nature Materials}}
  \textbf{\bibinfo{volume}{20}}, \bibinfo{pages}{341--345}
  (\bibinfo{year}{2021}).

\bibitem{Chauleau2020}
\bibinfo{author}{Chauleau, J.-Y.} \emph{et~al.}
\newblock \bibinfo{title}{{Electric and antiferromagnetic chiral textures at
  multiferroic domain walls}}.
\newblock \emph{\bibinfo{journal}{Nature Materials}}
  \textbf{\bibinfo{volume}{19}}, \bibinfo{pages}{386--390}
  (\bibinfo{year}{2020}).

\bibitem{Garcia2014}
\bibinfo{author}{Garcia, V.} \& \bibinfo{author}{Bibes, M.}
\newblock \bibinfo{title}{{Ferroelectric tunnel junctions for information
  storage and processing}}.
\newblock \emph{\bibinfo{journal}{Nature Communications}}
  \textbf{\bibinfo{volume}{5}}, \bibinfo{pages}{4289} (\bibinfo{year}{2014}).

\bibitem{Mehta1973}
\bibinfo{author}{Mehta, R.~R.}, \bibinfo{author}{Silverman, B.~D.} \&
  \bibinfo{author}{Jacobs, J.~T.}
\newblock \bibinfo{title}{{Depolarization Fields in Thin Ferroelectric Films}}.
\newblock \emph{\bibinfo{journal}{Journal of Applied Physics}}
  \textbf{\bibinfo{volume}{44}}, \bibinfo{pages}{3379--3385}
  (\bibinfo{year}{1973}).

\bibitem{DeLuca2017a}
\bibinfo{author}{{De Luca}, G.} \emph{et~al.}
\newblock \bibinfo{title}{{Nanoscale design of polarization in ultrathin
  ferroelectric heterostructures}}.
\newblock \emph{\bibinfo{journal}{Nature Communications}}
  \textbf{\bibinfo{volume}{8}}, \bibinfo{pages}{1419} (\bibinfo{year}{2017}).

\bibitem{Strkalj2020}
\bibinfo{author}{Strkalj, N.} \emph{et~al.}
\newblock \bibinfo{title}{{In-situ monitoring of interface proximity effects in
  ultrathin ferroelectrics}}.
\newblock \emph{\bibinfo{journal}{Nature Communications}}
  \textbf{\bibinfo{volume}{11}}, \bibinfo{pages}{5815} (\bibinfo{year}{2020}).

\bibitem{Puggioni2018a}
\bibinfo{author}{Puggioni, D.}, \bibinfo{author}{Giovannetti, G.} \&
  \bibinfo{author}{Rondinelli, J.~M.}
\newblock \bibinfo{title}{{Polar metals as electrodes to suppress the
  critical-thickness limit in ferroelectric nanocapacitors}}.
\newblock \emph{\bibinfo{journal}{Journal of Applied Physics}}
  \textbf{\bibinfo{volume}{124}}, \bibinfo{pages}{174102}
  (\bibinfo{year}{2018}).

\bibitem{Keeney2020}
\bibinfo{author}{Keeney, L.} \emph{et~al.}
\newblock \bibinfo{title}{{Ferroelectric Behavior in Exfoliated 2D Aurivillius
  Oxide Flakes of Sub-Unit Cell Thickness}}.
\newblock \emph{\bibinfo{journal}{Advanced Electronic Materials}}
  \textbf{\bibinfo{volume}{6}}, \bibinfo{pages}{1901264}
  (\bibinfo{year}{2020}).

\bibitem{Gradauskaite2020a}
\bibinfo{author}{Gradauskaite, E.} \emph{et~al.}
\newblock \bibinfo{title}{{Robust In-Plane Ferroelectricity in Ultrathin
  Epitaxial Aurivillius Films}}.
\newblock \emph{\bibinfo{journal}{Advanced Materials Interfaces}}
  \textbf{\bibinfo{volume}{7}}, \bibinfo{pages}{2000202}
  (\bibinfo{year}{2020}).

\bibitem{aurivilliusmain}
\bibinfo{author}{Aurivillius, B.}
\newblock \bibinfo{title}{{Mixed Bismuth Oxides with Layer Lattices. 1. The
  Structure Type of
  CaNb\textsubscript{2}Bi\textsubscript{2}O\textsubscript{9}}}.
\newblock \emph{\bibinfo{journal}{Arkiv. Kemi.}} \textbf{\bibinfo{volume}{1}},
  \bibinfo{pages}{463--480} (\bibinfo{year}{1949}).

\bibitem{Gradauskaite2021a}
\bibinfo{author}{Gradauskaite, E.}, \bibinfo{author}{Gray, N.},
  \bibinfo{author}{Campanini, M.}, \bibinfo{author}{Rossell, M.~D.} \&
  \bibinfo{author}{Trassin, M.}
\newblock \bibinfo{title}{{Nanoscale Design of High-Quality Epitaxial
  Aurivillius Thin Films}}.
\newblock \emph{\bibinfo{journal}{Chemistry of Materials}}
  \textbf{\bibinfo{volume}{33}}, \bibinfo{pages}{9439--9446}
  (\bibinfo{year}{2021}).

\bibitem{Spaldin2021}
\bibinfo{author}{Spaldin, N.~A.}, \bibinfo{author}{Efe, I.},
  \bibinfo{author}{Rossell, M.~D.} \& \bibinfo{author}{Gattinoni, C.}
\newblock \bibinfo{title}{{Layer and spontaneous polarizations in perovskite
  oxides and their interplay in multiferroic bismuth ferrite}}.
\newblock \emph{\bibinfo{journal}{The Journal of Chemical Physics}}
  \textbf{\bibinfo{volume}{154}}, \bibinfo{pages}{154702}
  (\bibinfo{year}{2021}).

\bibitem{Keeney2020a}
\bibinfo{author}{Keeney, L.} \emph{et~al.}
\newblock \bibinfo{title}{{Persistence of Ferroelectricity Close to Unit-Cell
  Thickness in Structurally Disordered Aurivillius Phases}}.
\newblock \emph{\bibinfo{journal}{Chemistry of Materials}}
  \textbf{\bibinfo{volume}{32}}, \bibinfo{pages}{10511--10523}
  (\bibinfo{year}{2020}).

\bibitem{Kittel}
\bibinfo{author}{Kittel, C.}
\newblock \emph{\bibinfo{title}{{Introduction to Solid State Physics}}}
  (\bibinfo{publisher}{John Wiley and Sonds, Inc}, \bibinfo{year}{2005}),
  \bibinfo{edition}{8th ed.} edn.

\bibitem{Steffes2019}
\bibinfo{author}{Steffes, J.~J.}, \bibinfo{author}{Ristau, R.~A.},
  \bibinfo{author}{Ramesh, R.} \& \bibinfo{author}{Huey, B.~D.}
\newblock \bibinfo{title}{{Thickness scaling of ferroelectricity in
  BiFeO\textsubscript{3} by tomographic atomic force microscopy}}.
\newblock \emph{\bibinfo{journal}{Proceedings of the National Academy of
  Sciences of the United States of America}} \textbf{\bibinfo{volume}{116}},
  \bibinfo{pages}{2413--2418} (\bibinfo{year}{2019}).

\bibitem{Streiffer1998}
\bibinfo{author}{Streiffer, S.~K.} \emph{et~al.}
\newblock \bibinfo{title}{{Domain patterns in epitaxial rhombohedral
  ferroelectric films. I. Geometry and experiments}}.
\newblock \emph{\bibinfo{journal}{Journal of Applied Physics}}
  \textbf{\bibinfo{volume}{83}}, \bibinfo{pages}{2742--2753}
  (\bibinfo{year}{1998}).

\bibitem{Emori2013}
\bibinfo{author}{Emori, S.}, \bibinfo{author}{Bauer, U.}, \bibinfo{author}{Ahn,
  S.~M.}, \bibinfo{author}{Martinez, E.} \& \bibinfo{author}{Beach, G.~S.}
\newblock \bibinfo{title}{{Current-driven dynamics of chiral ferromagnetic
  domain walls}}.
\newblock \emph{\bibinfo{journal}{Nature Materials}}
  \textbf{\bibinfo{volume}{12}}, \bibinfo{pages}{611--616}
  (\bibinfo{year}{2013}).

\bibitem{Velez2019}
\bibinfo{author}{V{\'{e}}lez, S.} \emph{et~al.}
\newblock \bibinfo{title}{{High-speed domain wall racetracks in a magnetic
  insulator}}.
\newblock \emph{\bibinfo{journal}{Nature Communications}}
  \textbf{\bibinfo{volume}{10}}, \bibinfo{pages}{4750} (\bibinfo{year}{2019}).

\bibitem{Dzyaloshinskii1964}
\bibinfo{author}{Dzyaloshinskii, I.~E.}
\newblock \bibinfo{title}{{Theory of Helicoidal Structures in Antiferromagnets.
  I. Nonmetals}}.
\newblock \emph{\bibinfo{journal}{Jetp}} \textbf{\bibinfo{volume}{19}},
  \bibinfo{pages}{960--971} (\bibinfo{year}{1964}).

\bibitem{Bogdanov2002}
\bibinfo{author}{Bogdanov, A.~N.}, \bibinfo{author}{R{\"{o}}{\ss}ler, U.~K.},
  \bibinfo{author}{Wolf, M.} \& \bibinfo{author}{M{\"{u}}ller, K.~H.}
\newblock \bibinfo{title}{{Magnetic structures and reorientation transitions in
  noncentrosymmetric uniaxial antiferromagnets}}.
\newblock \emph{\bibinfo{journal}{Physical Review B - Condensed Matter and
  Materials Physics}} \textbf{\bibinfo{volume}{66}}, \bibinfo{pages}{214410}
  (\bibinfo{year}{2002}).

\bibitem{LandauLifshitz}
\bibinfo{author}{Landau, L.~D.} \& \bibinfo{author}{Lifshitz, E.~M.}
\newblock \emph{\bibinfo{title}{{Statistical Physics: Volume 5}}}
  (\bibinfo{publisher}{Pergamon Press}, \bibinfo{address}{Oxford},
  \bibinfo{year}{1969}), \bibinfo{edition}{2nd} edn.

\bibitem{Erb2020}
\bibinfo{author}{Erb, K.~C.} \& \bibinfo{author}{Hlinka, J.}
\newblock \bibinfo{title}{{Vector, bidirector, and Bloch skyrmion phases
  induced by structural crystallographic symmetry breaking}}.
\newblock \emph{\bibinfo{journal}{Physical Review B}}
  \textbf{\bibinfo{volume}{102}}, \bibinfo{pages}{024110}
  (\bibinfo{year}{2020}).

\bibitem{Fusil2022}
\bibinfo{author}{Fusil, S.} \emph{et~al.}
\newblock \bibinfo{title}{{Polar Chirality in BiFeO\textsubscript{3} Emerging
  from A Peculiar Domain Wall Sequence}}.
\newblock \emph{\bibinfo{journal}{Advanced Electronic Materials}}
  \textbf{\bibinfo{volume}{2101155}}, \bibinfo{pages}{2101155}
  (\bibinfo{year}{2022}).

\bibitem{Song2020}
\bibinfo{author}{Song, J.}, \bibinfo{author}{Zhou, Y.},
  \bibinfo{author}{Padture, N.~P.} \& \bibinfo{author}{Huey, B.~D.}
\newblock \bibinfo{title}{{Anomalous 3D nanoscale photoconduction in hybrid
  perovskite semiconductors revealed by tomographic atomic force microscopy}}.
\newblock \emph{\bibinfo{journal}{Nature Communications}}
  \textbf{\bibinfo{volume}{11}}, \bibinfo{pages}{3308} (\bibinfo{year}{2020}).

\bibitem{Yankovich2014}
\bibinfo{author}{Yankovich, A.~B.} \emph{et~al.}
\newblock \bibinfo{title}{{Picometre-precision analysis of scanning
  transmission electron microscopy images of platinum nanocatalysts}}.
\newblock \emph{\bibinfo{journal}{Nature Communications}}
  \textbf{\bibinfo{volume}{5}}, \bibinfo{pages}{4155} (\bibinfo{year}{2014}).

\bibitem{Campanini2018}
\bibinfo{author}{Campanini, M.}, \bibinfo{author}{Erni, R.},
  \bibinfo{author}{Yang, C.~H.}, \bibinfo{author}{Ramesh, R.} \&
  \bibinfo{author}{Rossell, M.~D.}
\newblock \bibinfo{title}{{Periodic Giant Polarization Gradients in Doped
  BiFeO\textsubscript{3} Thin Films}}.
\newblock \emph{\bibinfo{journal}{Nano Letters}} \textbf{\bibinfo{volume}{18}},
  \bibinfo{pages}{717--724} (\bibinfo{year}{2018}).

\end{thebibliography}

\section*{\large{Acknowledgements}} E.G. and M.T. acknowledge the Swiss National Science Foundation under Project No. 200021\_188414. M.T. and M.F. acknowledge support by the EU European Research Council under Advanced Grant Program No. 694955-INSEETO. Q.N.M. acknowledges support by the Swiss National Science Foundation under project No. P2EZP2\_191872. M.C. and M.D.R. acknowledge support by the Swiss National Science Foundation under
Project No. 200021\_175926. M.F. acknowledges support by the Swiss National Science Foundation under Project No. 200021\_178825. 

\section*{\large{Author Contributions}} All authors discussed the results. E.G. and M.T. wrote the manuscript with M.F.. E.G. performed the thin-film growth, the ISHG measurements and the structural analysis. Scanning probe microscopy measurements were done by E.G. together with N.G.. Q.N.M. performed the phase field simulations. M.C. and M.D.R. carried out the STEM investigations. T.M. and B.D.H. performed tomographic PFM study. M.T. designed the experiment with E.G. and supervised the work jointly with M.F.

\section*{\large{Additional Information}}
\subsection{Competing Interests} The authors declare that they have no competing financial interests.

\subsection{Supplementary information} is available below.
\subsection{Correspondence and requests for materials} should be addressed to E.G. and M.T.

\clearpage


\section*{Supplementary material}

# Supplementary Information

**Supplementary Note 1. Epitaxial matching between perovskite oxides (BTO, BFO), Aurivillius compound BFTO and orthorhombic NGO (001)$_o$ substrate.**

|  | Lattice parameters (Å) | | | Mismatch (%) | |
|---|---|---|---|---|---|
|  | $a$ | $b$ | $\sqrt{a^2+b^2}$ | [100]$_{NGO}$ | [010]$_{NGO}$ |
| NGO[1] | 5,428 | 5,498 | | | |
| BTO[2] | 4,000 | 4,000 | 5,657 | -4,05 | -2,81 |
| BFO[3] | 3,965 | 3,965 | 5,607 | -3,20 | -1,95 |
| BFTO[4] | 5,470 | 5,439 | | -0,21 | 0,51 |

**Fig. S1.1 | Lattice parameters and the resulting lattice mismatch on NGO substrate.** Epitaxy between 001-oriented pseudocubic unit cell (denoted pc) and orthorhombic unit cell (denoted o) is achieved when $a_{pc}$ and $b_{pc}$ are kept equal, while diagonals of the unit cell are matched to $a_o$ and $b_o$[5]. This leads to the epitaxial relationship with [100]$_{pc}$ ∥ [1$\bar{1}$0]$_o$ and [010]$_{pc}$ ∥ [110]$_o$.

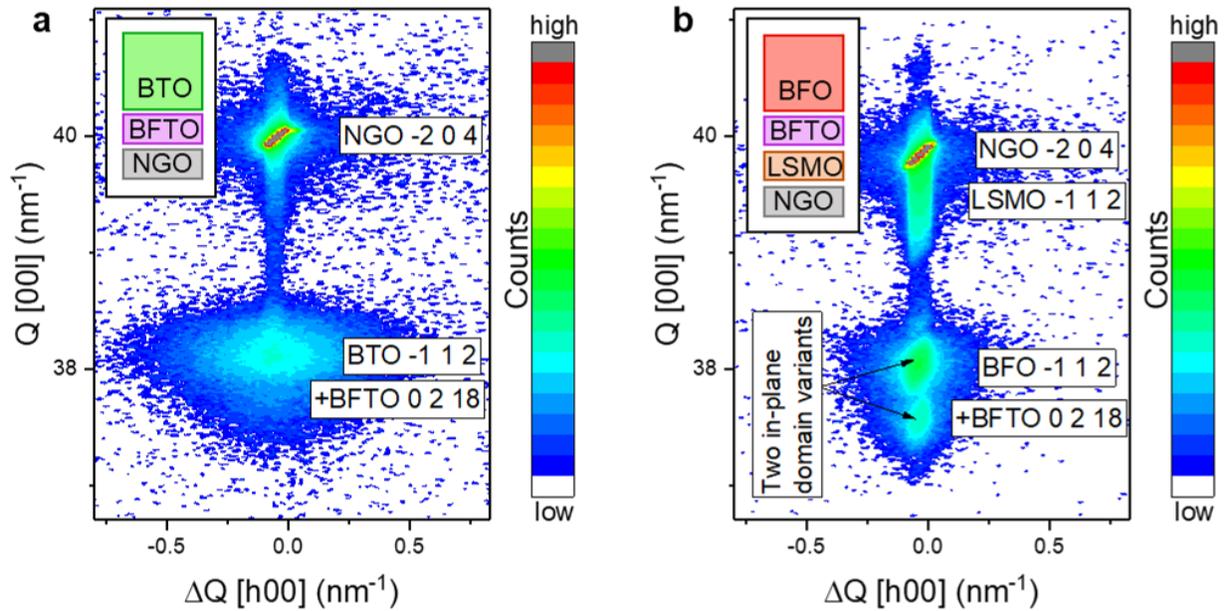

**Fig. S1.2 | Reciprocal space maps (out-of-plane Q [00l] and in-plane ΔQ[k00]) around the NGO (-2 0 4) reflection revealing the strain state of the heterostructures sketched. a**, Partial BTO relaxation in BTO|BFTO|NGO **b**, Coherently strained BFO|BFTO|LSMO|NGO. Split BFO (-1 1 2) reflection corresponds to two-in plane domain variants[6].

**Supplementary Note 2. ISHG monitoring of BTO grown on NGO (001)$_o$ and STO (001) buffered with 15 u.c. LSMO**

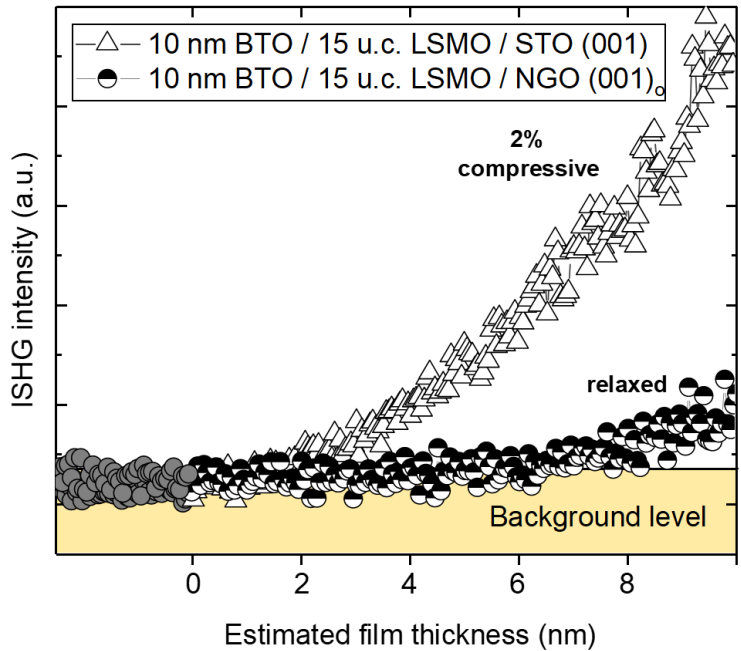

**Fig. S2 | ISHG signal tracking the BTO thin-film polarization during the deposition on different substrates:** on STO (001) (white triangle symbols) and NGO (001)$_o$ (white and black round symbols).

Macroscopic strain-relaxation in BTO films could cause the reduction of ferroelectric transition temperature towards or even below the growth temperature and, hence, prompt a reduced spontaneous ferroelectric polarization[7,8]. Despite the identical, partially relaxed strain-state for BTO films grown LSMO-buffered NGO and on BFTO-buffered NGO substrates, a clear onset of ferroelectricity from the first deposited u.c. is only observed in the latter case, highlighting the role of the in-plane polarized buffer. We show in addition that fully strained BTO films gown on LSMO-buffered STO substrates do exhibit a critical thickness. We thus exclude such strain-related mechanism as a reason for the delayed onset of ISHG signal in the film grown on LSMO buffer.

**Supplementary Note 3. AFM tomography of 5 u.c.-thick (approx. 25 nm) BFTO film.**

Figure S3.1 displays a montage of piezoresponse maps every 1 nm throughout the thickness of the Tomographic AFM dataset. A gradual transition is clearly apparent from well-ordered, aligned

domain stripes near the substrate to a high density of columnar domains near the surface. The completely dark regions in the deepest few frames represent depths that were not accessed across the entire field of view, and hence have no data to contribute. Multiple tomographs were acquired from the same specimen, with equivalent results as those in Figure S3.1.

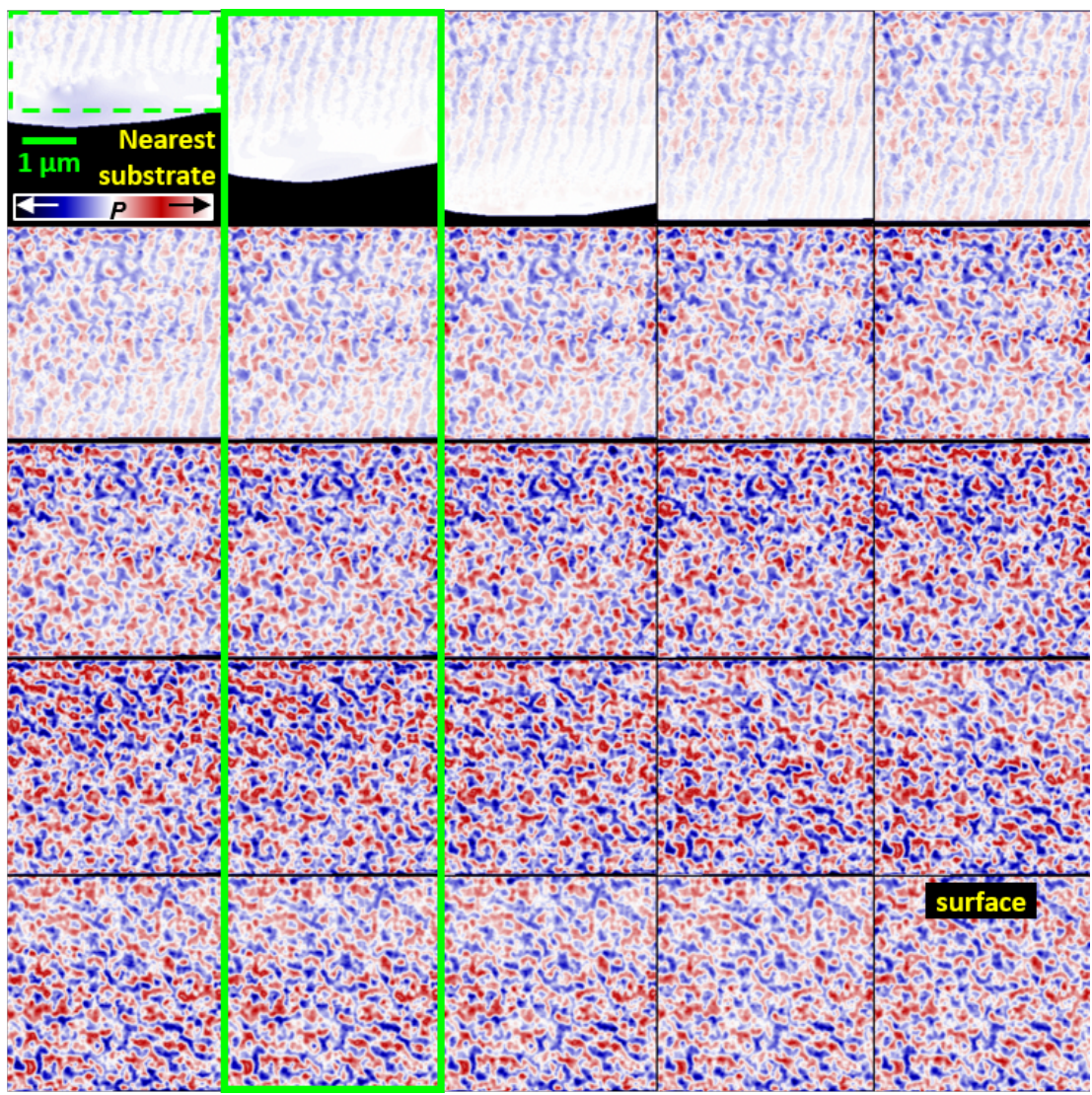

**Fig. S3.1 | Piezoresponse maps (4 μm × 4 μm) collected throughout the thickness of a 5 u.c.-thick layer of BFTO.** The piezoresponse maps were extracted from the Tomographic AFM dataset at every 1 nm of the BFTO film (approx. 25 nm) (arranged from nearest the substrate towards the surface).

To quantitatively assess the domain periodicity and alignment as a function of depth, 2D Fast Fourier Transforms (FFT) were computed in the top ∼45% of the field of view (dashed region) for each tomogram slice. Representative FFT results (from within the highlighted column) are shown in Figure S3.2a, revealing the magnitude (color) and any directionality (Cartesian with respect to the center as indicated) of periodicities ranging over a full field of view ± 0.02/nm. A weak but identifiable peak is apparent at ∼300 nm and -13.6° for the initial unit cells of the thin film (dotted overlays). As the film thickness increases, however, these peaks become difficult to distinguish from the apparently less ordered columnar domains.

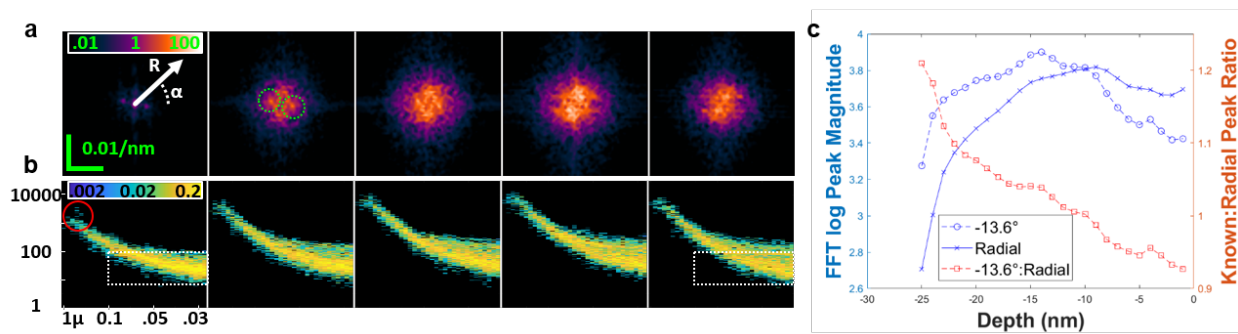

**Fig. S3.2 | 2D FFT computed for each tomogram slice shown in Fig. S3.1.** (a,b) Representative FFT results. (c) 2D FFT magnitude for every 1 nm deeper into the sample, specifically for two distinct periodicities: i) the peak initial domain pattern (∼300 nm and -13.6°); and ii) the overall peak radial periodicity. The ratio of these signals is also determined (right axis).

To analyze any orientationally-independent periodicities, the 2D FFT data per depth is furthermore recast in SI Figure Figure S3.2b — again for the representative frames from (a), though in fact calculated at every depth. These multidimensional histograms represent the range of detected FFT magnitudes (y axis) for distinct wavelengths (x-axis, instead of frequencies and hence shown as a log scale). The color contrast thus indicates the percent of data points with any given

pair of spatial periodicity and magnitude. The peak magnitude nearest ∼300 nm (solid circular overlay) is obvious for the deepest frames, but diminishingly dominates as film thickness increases because other longer and shorter wavelength periodicities develop. Comparing the contrast within the spatially identical dotted rectangular overlays is especially revealing, confirming the increased prevalence of domain periodicities on the order of 100 down to 30 nm as film thickness increases. The resulting high density of conducting domain walls can correspondingly serve as a 'phantom' back electrode.

SI Figure Figure S3.2c displays the 2D FFT magnitude for every 1 nm deeper into the sample (i.e. from every tomogram slice in SI Figure Figure S3.1), specifically for two distinct periodicities: i) the peak initial domain pattern (∼300 nm and -13.6°); and ii) the overall peak radial periodicity. The ratio of these signals is also determined (right axis), with a crossover at a depth of -10 nm indicative of the transition from aligned and regularly repeating stripe domains to finer columnar domains at a thickness of ∼15 nm (3-4 unit cells).

**Supplementary Note 4. Continuity of polar displacements at the BFO|BFTO interface.**

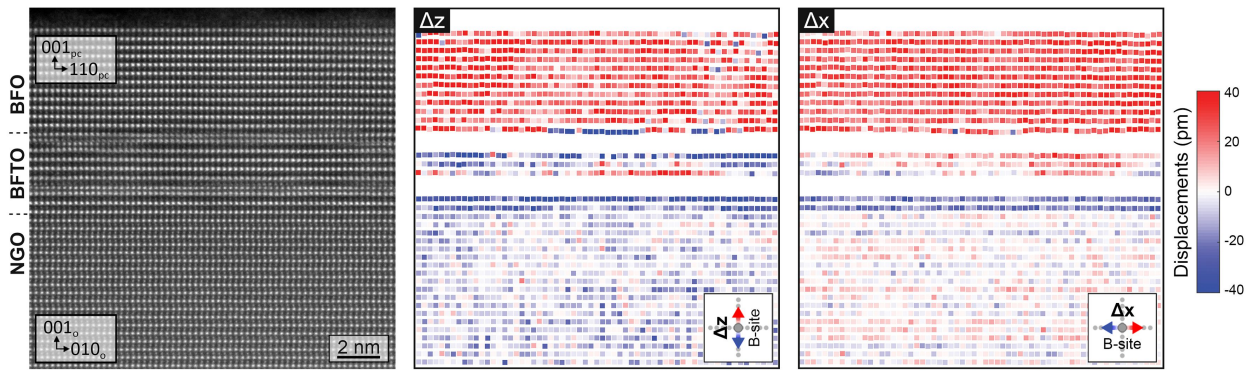

**Fig. S4 | Analysis of the in-plane and out-of-plane displacements in the BFO|BFTO|NGO heterostructure.** The polar displacements of the B-site cations atomic columns from the center of their two nearest A-site neighbors were measured from the atomically resolved HAADF-STEM image (left). Right: displacement maps along the out-of-plane ($\Delta z$) and in-plane ($\Delta x$) directions as shown by the inserts.

## Supplementary Note 5. Uniaxial in-plane polarization and domain configuration in BFO films grown on BFTO (1 u.c.)|LSMO (20 u.c.)|NGO (001)$_o$

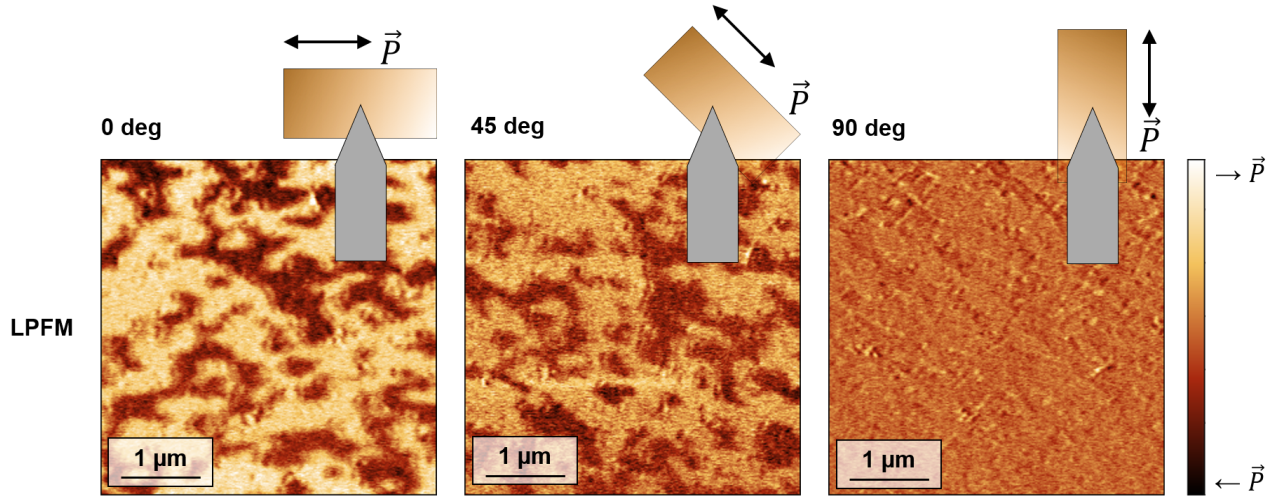

**Fig. S5.1 | Uniaxial in-plane polarization in BFO determined with rotation-dependent PFM study.** The uniaxial nature of in-plane polarization in BFO films on BFTO|LSMO|NGO was confirmed using rotation-dependent PFM via deconvolution of the cantilever torsion and buckling modes. With the cantilever aligned with the [100]$_o$ NGO axis, the uniaxial polarization of BFO and BFTO below are at right angles to the cantilever, meaning that the piezoresponse is now predominantly recorded as the lateral PFM signal due to the torsion modes. At 45° sample rotation the domains are measured in both LPFM (torsion) and VPFM (buckling) channels, hence the LPFM signal weakens. With the sample rotated by 90° and the cantilever aligned with the [010]$_o$ NGO axis (a), the signal is now no longer in the LPFM channel and is instead recorded in the vertical PFM channel as the buckling of the cantilever.

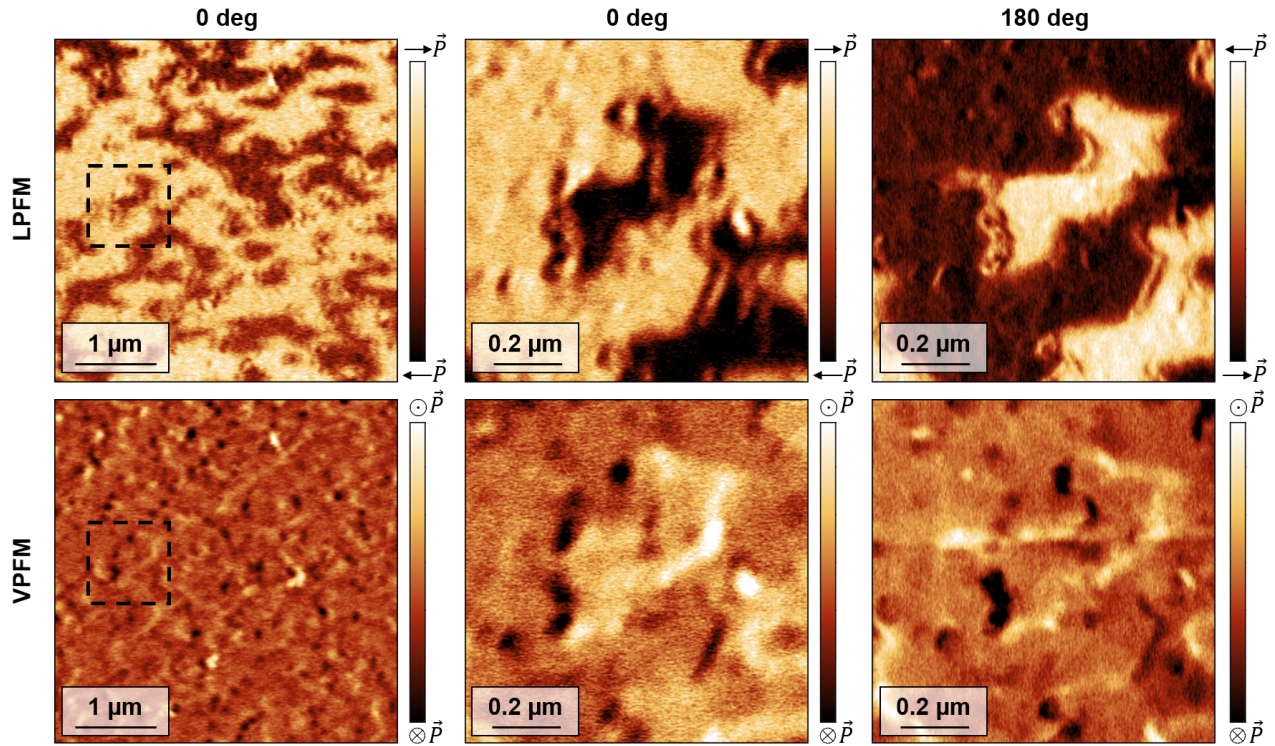

**Fig. S5.2 | Confirmation of out-of-plane polarization originating at the charged sections of domain walls via rotation-dependent PFM** The VPFM signal appearing at the charged domain walls could have different origins: it could arise from the buckling due to additional in-plane polarization components at the walls (perpendicular to the uniaxial in-plane polarization axis) or due to the deflection force from the pure out-of-plane polarization. In order to distinguish which of the two is the source of the observed behaviour, we select a micron-sized area (marked in larger LPFM and VPFM scans) to resolve the domain wall signal. Then the sample is rotated by 180° and the same area is rescanned. The LPFM signal gets inverted, but the VPFM signal at the walls does not. This confirms that the signal has its origin as cantilever deflection and hence implies pure out-of-plane polarization localized at the walls.

**Supplementary Note 6. Absence of Néel-type domain walls and homochirality in BFO films grown on LSMO (20 u.c.)|NGO (001)$_o$**

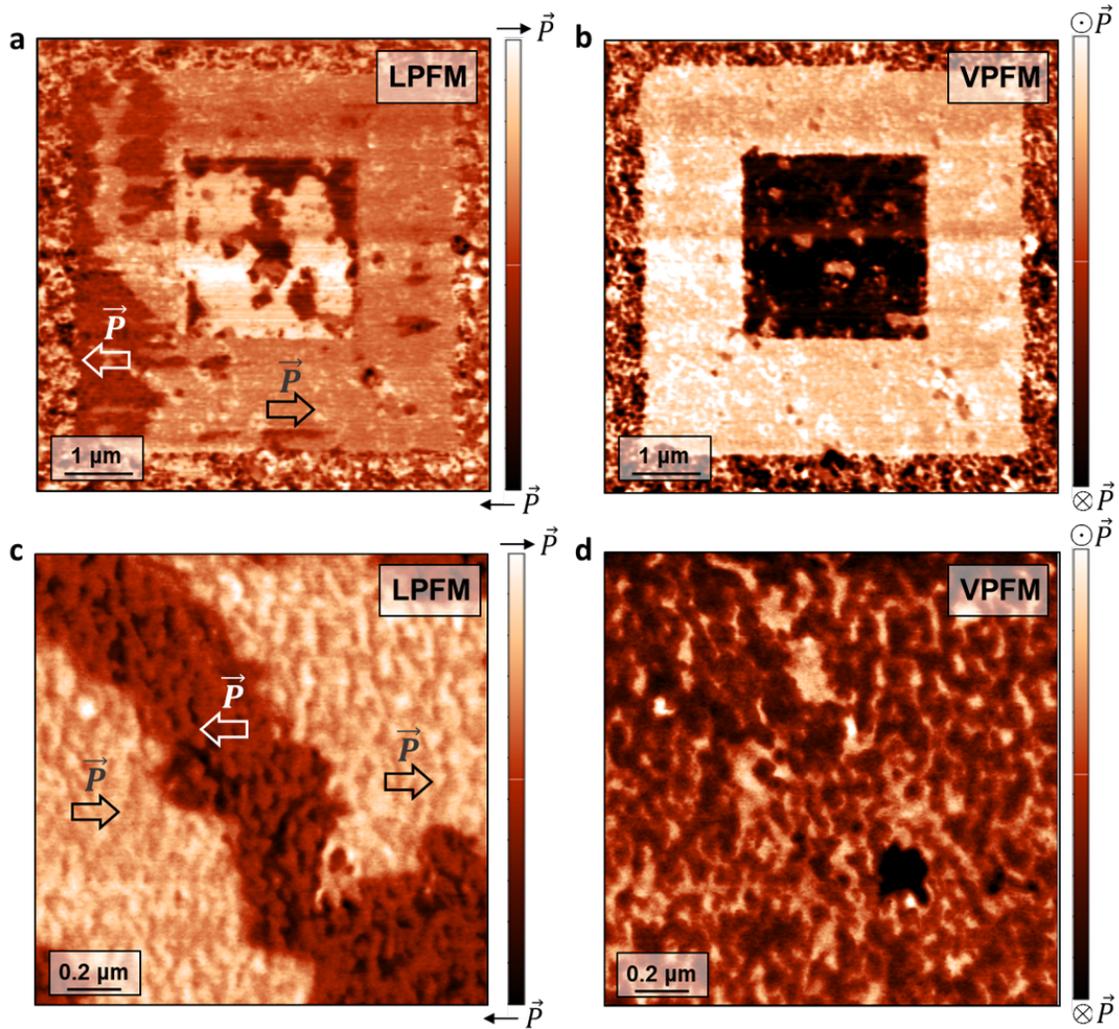

**Fig. S6 | PFM study of switched ferroelectric domains in BFO|LSMO|NGO confirming absence of domain wall topology.** LPFM (a) and (b) VPFM scans of the BFO film on LSMO|NGO. The pristine BFO film exhibits 180° out-of-plane-polarized nanodomain formation (visible outside of the reversibly poled area). No stabilization of charged domain walls is detected. Larger in-plane domains can be temporarily stabilized upon the local electrical poling. We confirm that these domains exhibit uniaxial in-plane polarization similarly to Fig. S3.1. Higher-magnification LPFM (c) and VPFM (d) scans of artificially-formed HH and TT walls show no signature of Néel-type chiral domain walls in this film, confirming that the BFTO buffer layer brings around the symmetry-breaking required for the net chirality in the BFO domain walls to emerge.

29

**Supplementary note 7: Phase field simulation of homochiral domain walls**

The chirality of a domain wall can be judged by evaluating a Lifschitz invariant term $P_x \nabla P_z - \nabla P_x P_z$ [9]. In most ferroelectric materials that host Néel-type walls, left-handed and right-handed walls are equally common. However, due to structural Dzyaloshinskii-Moriya interactions[10] one chirality can become favored over the other, for which the above-mentioned term enters the free energy expansion. We start by creating a minmal model free energy for the polarization in cubic BiFeO$_3$, which has the free energy expansion:

$$F = \frac{a}{2}(P_x^2 + P_y^2 + P_z^2)^2 + \frac{b}{4}(P_x^2 + P_y^2 + P_z^2) + \frac{c}{4}(P_x^4 + P_y^4 + P_z^4) \qquad (1)$$

Where $x, y, z$ correspond to the pseudocubic axes. From our experiments we know, that the only polarization axes that are observed are $[110]_{\text{pc}}$ and $[001]_{\text{pc}}$. We thus perform a 45° rotation in the x-y-plane of our coordinate system, in order to align $P_x$ with $[110]_{\text{pc}}$ and $P_y$ with $[1\bar{1}0]_{\text{pc}}$. Since the latter polarization direction is not observed, we then set it to zero, allowing us to reduce the phase space to the polarization directions $P_x$ ($[110]_{\text{pc}}$) and $P_z$ ($[001]_{\text{pc}}$). The new free energy expression reads:

$$F = \frac{a}{2}(P_x^2 + P_z^2) + \frac{b}{4}(P_x^2 + P_z^2)^2 + \frac{c}{8}(P_x^4) + \frac{c}{4}P_z^4 \qquad (2)$$

$$+ \sigma(P_x \nabla P_z - \nabla P_x P_z) \qquad (3)$$

$$- E_x P_x - E_z P_z \qquad (4)$$

where $a$, $b$, $c$ are the Landau expansion parameters of the polar mode. We included the Lifshitz invariant with amplitude $\sigma$ and the last two terms describe the coupling to an electric field. We use the Landau parameters shown in Table 1, to qualitatively reproduce the behavior of the BFO in the BFO|BFTO heterostructure. These parameters reproduce an absolute polarization of $P_x = \sqrt{(2)}P_z$ at the 4 global minima at $[111]_{pc}$, $[\bar{1}\bar{1}1]_{pc}$, $[11\bar{1}]_{pc}$ or $[\bar{1}\bar{1}\bar{1}]_{pc}$, i.e. the uniaxial in-plane polarization in BFO we observe experimentally. In addition, we apply an electric field $E_z$ along $-z$ to simulate the field arising due to the interfacial bound charge. This reduces the number of stable polarization states from four to two, $[11\bar{1}]_{pc}$ and $[\bar{1}\bar{1}\bar{1}]_{pc}$ that are downwards polarized, in accordance with our experimental observations. We solve the Euler equations resulting from (4) using the PDE solvers implemented in the FiPy package[11]. The resulting phase field simulation shown in Figure S6 reproduces our experimental PFM data and confirms that the homochirality of domain walls in the BFO films grown on BFTO can be emulated by considering a ferroelectric DMI-like interaction[10,12].

| $a$ | $b$ | $c$ | $\sigma$ |
|---|---|---|---|
| -1. | 2. | 2. | 0.35 |

Table 1: Landau parameters used for the simulation

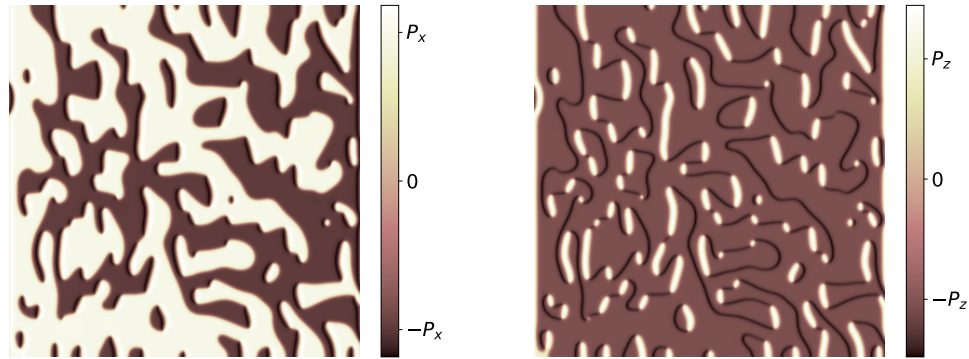

**Fig. S7 | Phase field simulation of chiral domain walls in systems with non-zero Lifshitz invariant.** The simulated in-plane-polarized domains and the out-of-plane polarization emergence at the domain walls are comparable to the measured LPFM and VPFM scans of BFO|BFTO system, respectively (see main text Fig. 4e,f.)

# References

1. Vasylechko, L., Akselrud, L., Morgenroth, W., Bismayer, U., Matkovskii, A. & Savytskii, D. Crystal structure of NdGaO$_3$ at 100 K and 293 K based on synchrotron data. *Journal of Alloys and Compounds* **297**, 46–52 (2000).

2. Buttner, R. H. & Maslen, E. N. Structural parameters and electron difference density in BaTiO$_3$. *Acta Crystallographica Section B* **48**, 764–769 (1992).

3. Kubel, F. & Schmid, H. Structure of a ferroelectric and ferroelastic monodomain crystal of the perovskite BiFeO$_3$. *Acta Crystallographica Section B* **46**, 698–702 (1990).

4. Hervoches, C. H., Snedden, A., Riggs, R., Kilcoyne, S. H., Manuel, P. & Lightfoot, P. Structural behavior of the four-layer aurivillius-phase ferroelectrics SrBi$_4$Ti$_4$O$_{15}$ and Bi$_5$Ti$_3$FeO$_{15}$. *Journal of Solid State Chemistry* **164**, 280–291 (2002).

5. Boschker, H., Mathews, M., Houwman, E. P., Nishikawa, H., Vailionis, A., Koster, G., Rijnders, G. & Blank, D. H. Strong uniaxial in-plane magnetic anisotropy of (001)- and (011)-oriented $La_{0.67}Sr_{0.33}MnO_3$ thin films on $NdGaO_3$ substrates. *Physical Review B - Condensed Matter and Materials Physics* **79**, 214425 (2009).

6. Chen, Z. H., Damodaran, A. R., Xu, R., Lee, S. & Martin, L. W. Effect of "symmetry mismatch" on the domain structure of rhombohedral $BiFeO_3$ thin films. *Applied Physics Letters* **104**, 182908 (2014).

7. Damodaran, A. R., Breckenfeld, E., Chen, Z., Lee, S. & Martin, L. W. Enhancement of Ferroelectric Curie Temperature in $BaTiO_3$ Films via Strain-Induced Defect Dipole Alignment. *Advanced Materials* **26**, 6341–6347 (2014).

8. Strkalj, N., De Luca, G., Campanini, M., Pal, S., Schaab, J., Gattinoni, C., Spaldin, N. A., Rossell, M. D., Fiebig, M. & Trassin, M. Depolarizing-Field Effects in Epitaxial Capacitor Heterostructures. *Physical Review Letters* **123**, 147601 (2019).

9. Houchmandzadeh, B., Lajzerowicz, J. & Salje, E. Order parameter coupling and chirality of domain walls. *Journal of Physics: Condensed Matter* **3**, 5163–5169 (1991).

10. Zhao, H. J., Chen, P., Prosandeev, S., Artyukhin, S. & Bellaiche, L. Dzyaloshinskii–Moriya-like interaction in ferroelectrics and antiferroelectrics. *Nature Materials* **20**, 341–345 (2021).

11. Guyer, J. E., Wheeler, D. & Warren, J. A. Fipy: Partial differential equations with python. *Computing in Science Engineering* **11**, 6–15 (2009).

12. Erb, K. C. & Hlinka, J. Vector, bidirector, and Bloch skyrmion phases induced by structural crystallographic symmetry breaking. *Physical Review B* **102**, 024110 (2020).



\end{document}